\documentclass[onecolumn,prl]{revtex4}
\usepackage{bm}
\usepackage{epsfig}
\usepackage{amsmath}
\begin{document}
\title{Barrierless Electronic Relaxation in Solution: An analytically solvable model with arbitrary coupling}
\author{Aniruddha Chakraborty \\
School of Basic Sciences, Indian Institute of Technology Mandi,\\
Mandi, Himachal Pradesh, 175001, India}
\date{\today }
\begin{abstract}
\noindent  In our recent publication, we have proposed an analytical method for solving the problem of electronic relaxation in solution, modeled by a particle undergoing diffusive motion under the influence of two arbitrary potentials and the coupling between the two potentials was assumed to be localized at one point. In this paper we have extended our work to deal with the case of delocalized coupling between two potentials. The average and long term rate constant for delocalized coupling is presented.
\end{abstract}
\maketitle
Relaxation of an excited electronic potential energy surface of a molecule which is in a polar solvent is an interesting problem both theoretically and experimentally \cite{Harris1,Harris2,Harris3,Fleming,AniJCP}. A molecule immersed in a polar solvent can be put on an electronically excited potential energy surface by the absorption of electromagnetic radiation. The excited molecule executes a walk on that potential energy surface, which may be taken to be random as due to the presence of polar solvent. As the molecule moves it may undergo non-radiative decay from certain regions of that surface and also undergoes radiative decay from anywhere on that surface. In our earlier publication \cite{AniJCP}, we have proposed an analytical method for solving the problem of electronic relaxation in solution, modeled by a particle undergoing diffusive motion under the influence of two arbitrary potentials and the coupling between the two potentials was assumed to be represnted by a Dirac Delta function. In this paper we have extended our model to deal with the case where the coupling between two potentials can be taken as any arbitrary function. Again the problem is to calculate the probability that the molecule will still be in the electronically excited potential energy surface after a finite time $t$.\par
We denote  the probability that the molecule can survive on the electronically excited potential energy surface by $P_{e}(x,t)$. We also use $P_{g}(x,t)$ to represent the probability that the molecule would be found in the ground state potential energy surface. It is very usual to assume the motion on both the potentiale energy surface to be one dimensional and diffusive, the relevent coordinate being denoted by $x$. It is also common to assume that the motion on both the potential energy surface is overdamped. Thus both the probability $P_{e}(x,t)$ and $P_{g}(x,t)$ may be found at $x$ at the time $t$ obeys a modified Smoluchowskii equation.
\begin{eqnarray}
\frac{\partial P_e(x,t)}{\partial t} = {\cal L}_e P_e(x,t) - k_r P_e(x,t) - k_0 S(x) P_g(x,t)  \\ \nonumber
\frac{\partial P_g(x,t)}{\partial t} ={\cal L}_g P_g(x,t) - k_r P_g(x,t) - k_0 S(x) P_e(x,t). \nonumber
\end{eqnarray}
In the above 
\begin{equation}
{\cal L}_i= A \frac{\partial^2}{\partial x^2}+\frac{\partial}{\partial x} \frac{dV_i(x)}{dx}.
\end{equation}
$V_i(x)$ is the potential causing the drift of the particle, $S(x)$ is a position dependent sink function, $k_0$ is the rate of nonradiative decay and $k_r$ is the rate of radiative decay. We have taken $k_r$ to be independent of position. $A$ is the diffusion coefficient. Our model is more general than the model proposed by Sebastian et. al., \cite{Kls1,Kls2}, in the sense that our model considers the motion on the ground state potential energy surface explicitely. It is quite unlikely that the shape of the ground state potential energy surface do not play any role in the electronic relaxation process, but unfortunately this fact was not considered in any of the earlier studies \cite{Kls1,Kls2,Oxtoby,Marcus,Bagchi,Nishijima,Szabo}.
Before we excite, the molecule is in the ground state potential energy surface, and as the solvent is at a finite temperature, its distribution over the coordinate $x$ is random. From this it undergoes Franck-Condon excitation to the excited state potential energy surface. So, $x_0$ the initial position of the particle, on the excited state potential energy surface is random. We assume it to be given by the probability density $P^{0}_{e} (x_0)$. In the follwoing we provide a general procedure for finding the exact analytical solution of Eq. (1). The Laplace transform ${\cal P}_i(x,s)=\int_{0}^{\infty} P_i(x,t)e^{-st} dt$ obeys
\begin{eqnarray}
[s-{\cal L}_e+k_r] {\cal P}_e(x,s)+k_0 S(x) {\cal P}_g (x,s) = P^0_e(x_0) \\ \nonumber
[s-{\cal L}_g+k_r] {\cal P}_g(x,s)+k_0 S(x) {\cal P}_e (x,s) = 0, \nonumber
\end{eqnarray}
where $P^0_e(x_0)=P_e(x,0)$ is the initial distribution at the electronically excited state and $P_g(x,0)=0$ is the initial distribution at the electronically ground state. 
\begin{equation}
 \left(
\begin{array}{c}
{\cal P}_e (x,s) \\
{\cal P}_g (x,s)
\end{array}
\right) = \left(
\begin{array}{cc}
s-{\cal L}_e+k_r & k_0 S(x) \\
k_0 S(x) & s-{\cal L}_g+k_r
\end{array}
\right)^{-1}
\left(
\begin{array}{c}
P^0_e(x) \\
0
\end{array}
\right)  ,
\end{equation}
Using the partition technique \cite{Lowdin}, solution of this equation can be expressed as 
\begin{equation}
{\cal P}_e(x,s)=\int_{-\infty}^{\infty} dx_0 G(x,s;x_0)P^0_e(x_0),
\end{equation}
where $G(x,s;x_0)$ is the Green's function defined by
\begin{equation}
G(x,s;x_0)=\left < x \left|[s-{\cal L}_e+ k_r - {k_0}^2 S[s-{\cal L}_g+k_r]^{-1}S]^{-1}\right| x_0 \right>
\end{equation}
The above equation is true for any general $S$. This expressions simplify considerably if S is a Dirac Delta function located at $x_c$ \cite{AniJCP}. But the majority of the problems of interest, however do not corresponds to a localized coupling and one requires different forms of coupling $S(x)$ for proper description of dynamics in different cases. We express the arbitrary coupling function $S(x)$ in terms of a linear combination of Dirac Delta functions \cite{SKG,Szabo}. Expressing the arbitrary coupling function $S(x)$ in terms of Dirac Delta functions has the advantage that it can be solved exactly by using analytical methods. An arbitrary coupling $S(x)$ can be written as 
\begin{equation}
S(x)=\int_{-\infty}^{\infty}dx'S(x')\delta(x-x')
\end{equation}
and the above integral can be discritized as
\begin{equation}
S(x)=\sum_{j=1}^{N}k_{j}\delta(x-x_j),
\end{equation}
here $k_j$ are constants, given by
\begin{equation}
k_{j}=w_{j}S(x_{j}).
\end{equation}
The weight factor $w_j$ varies depending on the scheme of discritization used \cite{SKG}.
In the operator notation $S(x)$ may be written as $V=\sum_{j=1}^{N}K_{j}S_{j}=\sum_{j=1}^{N}K_{j}|x_{j}\rangle \langle x_{j}|$. Then
\begin{equation}
G_{11}^{}(x,x_0;E )=\langle x|[s-{\cal L}_e+ k_r - \sum_{j=1}^{N}K_{j}^2G_g^0(x_j,x_j;s )S_{j}]^{-1}|x_0\rangle ,
\end{equation}
where
\begin{equation}
G_2^0(x,x_0;E )=\langle x|(s-{\cal L}_g+ k_r)^{-1}|x_0\rangle ,
\end{equation}
and corresponds to propagation of the particle starting at $x_0$ on the second potential, in the absence of coupling to the
first potential. Now we use the operator identity \cite{Sebastian,Sebas}
\begin{widetext}
\begin{equation}
\label{18} (s-{\cal L}_e+ k_r- \sum_{j=1}^{N} K_{j}^2G^0_g(x_j,x_j;s)S_{j})^{-1}=
(s-{\cal L}_e+ k_r)^{-1}+(s-{\cal L}_e+ k_r)^{-1} \sum_{j=1}^{N} K_{j}^2G_g^0(x_j,x_j;s)S_{j}[s-{\cal L}_g+ k_r - \sum_{j=1}^{N} K_{j}^2G_g^0(x_j,x_j;s )S_{j}]^{-1}.
\end{equation}
\end{widetext}
Inserting the resolution of identity $I=\int_{-\infty }^\infty dy$
$ |y\rangle $ $\langle y|$ in the second term of the above equation and integrating, we arrive at
\begin{align}
\label{20}G_{11}^{}(x,x_0;s)=&G_e^0(x,x_0;s)+ \sum_{j=1}^{N} K_{j}^2G_e^0(x,x_j;s) \times G_g^0(x_j,x_j;s)G_{11}(x_j,x_0;s).
\end{align}
\newline
Considering the above equation at the discrete points $x_{i}$, we obtain a set of linear equations, which can be written as 
\begin{align}
AP=Q,
\end{align}
where the elements of the matrices $A=[a_{ij}]$, $P=[p_i]$ and $Q=[q_i]$ are given by
\begin{align}
&a_{ij}=- k_{i}^2 G^{0}_{e}(x_i,x_j;s)G^{0}_{g}(x_j,x_j;s)+\delta_{ij}\\ \nonumber
&p_i=G_{11}(x_i,x_0;s)\\ \nonumber
&q_{i}= G^{0}_{e}(x_i,x_0;s)
\end{align}
One can solve the matrix equation {\it i.e.} Eq. (15) easily and obtain $G_{11}(x_i,x_0;E)$ for all $x_i$. Eq. (14) then yield $G_{11}(x,x_0;E)$. Using these expressions for the Green's function in Eq. (5) we can calculate ${\cal P}_e(x,s)$ explicitely. The expressions that we have obtained for ${\cal P}_e(x,s)$ are quite general and are valid for any potentials. 
Using this Green's function in Eq. (5) one can caluclate ${\cal P}_e(x,s)$ explicitely. Here we are interested to know the survival probability at the electronically excited state potential energy surface $ P_e(t) = \int_\infty^\infty dx P_e(x,t)$. It is possible to evaluate Laplace Transform  ${\cal P}_e(s)$ of $P_e(t)$ directly. The average rate constant can be defined as ${k_{I}}^{-1}=\int_0^{\infty}dt P_e(t)$ and also a long term rate constant $k_L$ as 
\begin{equation}
k_{L}= - \lim_{t \rightarrow \infty} \frac{d}{dt} \ln P_e(t).
\end{equation} 
Clearly ${k_I}^{-1}={\cal P}_e(x,0)$ and $k_L$ is the negative pole of ${\cal P}_e(x,0)$ closest to the origin.
The average rate constant $k_I$ is thus given by
\begin{equation}
{k_I}^{-1}=\lim_{s\rightarrow 0} (k_r+s)^{-1} \left[1 - \sum_{i=1}^{N}k_i{\cal P}_e(x_i,s) \right]
\end{equation}
From the above equation we can understand that ${\cal P}_e(s)$ depends on $G^0_g(x_c,s;x_c)$. The expression that we have obtained for ${\cal P}_e(s)$, $k_I$ and $k_L$ are quite general and are valid for any set of potentials.

\end{document}